**Full title:** Phase memory in a dissipative time crystal
**Short title:** DTC phase memory

**Authors**
Darmindra Arumugam

**Affiliations**
Jet Propulsion Laboratory, California Institute of Technology, Pasadena, 91109, California, USA

**Abstract**
Dissipative time crystals exhibit persistent collective oscillations whose temporal phase emerges spontaneously from nonlinear many-body dynamics. Whether this phase can survive interruption of the excitation sustaining temporal order, however, remains an open question. Here, phase memory is demonstrated in a room-temperature Rydberg dissipative time crystal through controlled, microsecond-scale interruption of its optical coupling field. Following brief interruptions, the recovered oscillations retain their previously established phase, demonstrating that spontaneously selected temporal order can preserve information about its dynamical history. Measurements of phase-continuation fidelity and phase-transfer relationships reveal a characteristic memory timescale of approximately 4.3 μs. Longer interruptions progressively suppress phase retention, leading to reproducible phase resetting rather than random recovery. Nonlinear driven–dissipative mean-field simulations reproduce the principal experimental features, while a reduced analytical description captures the crossover through competition between retained phase information and restoration-induced phase selection. These findings identify phase memory as a measurable property of dissipative time-crystalline order and provide a framework for investigating the persistence and recovery of temporal information in nonequilibrium many-body systems.

## INTRODUCTION

Spontaneous symmetry breaking provides a fundamental framework for understanding the emergence of order in physical systems. Although conventional crystals break spatial translation symmetry, the concept of a time crystal extends spontaneous ordering into the temporal domain, giving rise to periodic dynamics that break the time-translation symmetry of the underlying system [1,2]. Despite fundamental constraints prohibiting continuous time crystallinity in thermal equilibrium [3], nonequilibrium systems offer a route to establishing persistent temporal order. This possibility has motivated extensive theoretical investigations [4,5] and experimental realizations of discrete time crystals in periodically driven many-body systems [6–8].

Driven–dissipative systems provide a particularly rich framework for the emergence of temporal order, in which the interplay between coherent interactions, external excitation and dissipation can stabilize dynamical phases inaccessible in equilibrium [9–11]. Experimental realizations of dissipative time crystals (DTCs) have demonstrated persistent collective oscillations across diverse physical platforms, including atom–cavity systems, interacting rare-earth ions, semiconductor spin ensembles and Rydberg atomic gases [12–16]. In continuous time crystals, the oscillation period emerges from the internal dynamics rather than being imposed by an external periodic drive, while the spontaneous selection of the oscillation phase reflects the breaking of continuous time-translation symmetry. The robustness of these oscillations against temporal perturbations establishes an important characteristic of dissipative time-crystalline order.

Beyond the emergence and stability of temporal order, an important question concerns its dynamical controllability. Synchronization and entrainment are well-established characteristics of classical and quantum nonlinear oscillators [17,18], and recent investigations have revealed rich synchronization dynamics, transitions between distinct time-crystalline states and bifurcations in driven–dissipative many-body systems [19–22]. Injection locking of Rydberg dissipative time crystals has further demonstrated that spontaneously established temporal order can be externally controlled and stabilized [23]. Although phase resetting has been extensively investigated in nonlinear dynamical systems [24,25], the fate of spontaneously established temporal order following interruption of its sustaining excitation remains poorly understood. A central question is whether a dissipative time crystal can retain a memory of its previously selected phase and recover that phase when excitation is restored.

Interacting Rydberg atomic ensembles provide an attractive platform for investigating this question. Strong interatomic interactions, coherent optical excitation and dissipation give rise to nonlinear collective behavior, nonequilibrium phase transitions and spontaneously synchronized dynamics [16,19,26–28]. The optical accessibility of these systems permits rapid interruption of Rydberg excitation on timescales considerably shorter than the collective oscillation period. Such interruptions provide a means of probing the persistence of phase information within the evolving atomic state and distinguishing phase-coherent recovery from the selection of a preferred phase upon renewed excitation.

Here, phase memory is demonstrated in a DTC realized in a continuously driven, strongly interacting Rydberg atomic ensemble. By temporarily extinguishing the optical coupling field, collective oscillations are shown to recover their previously established phase following brief interruptions, demonstrating that information about spontaneously selected temporal order can persist despite removal of its sustaining excitation. The phase memory exhibits a characteristic timescale of approximately 4.3 μs, beyond which recovery progressively gives way to reproducible phase resetting. Nonlinear mean-field simulations reproduce the principal features of this behavior, while a reduced analytical description captures the competition between retained phase information and restoration-induced phase selection. Together, these findings identify phase memory as a measurable dynamical property of dissipative time-crystalline order.

## RESULTS

### Phase-coherent recovery following optical interruption

To test whether a self-sustained collective oscillation retains its temporal phase following interruption of the sustaining excitation, a dissipative time crystal is realized in a strongly interacting Rydberg atomic ensemble. Rydberg systems provide a particularly suitable platform for investigating this question because strong interatomic interactions support nonlinear collective dynamics, while optical excitation permits rapid and precisely controlled perturbations of the oscillatory state. The experiment employs a room-temperature rubidium vapor addressed by counter-propagating probe (780 nm) and coupling (480 nm) fields (Fig. 1a) in a ladder-type excitation scheme. As illustrated in Fig. 1b, the optical fields couple the $5S_{1/2}$ ground state to the $5P_{3/2}$ intermediate state and subsequently to the $63D_{5/2}$ Rydberg manifold, with corresponding Rabi frequencies $\Omega_p$ and $\Omega_c$. A static magnetic field of approximately 19 G, applied along the optical axis, lifts the degeneracy of the Rydberg Zeeman sublevels. The probe and coupling-field polarizations are independently optimized to enhance the DTC oscillation contrast, enabling competing excitation pathways whose interaction-induced nonlinearities, optical pumping and dissipation support self-sustained collective oscillations in probe transmission [16,19,23]. Once stable DTC oscillations are established, an acousto-optic modulator (AOM) temporarily extinguishes the coupling field ($\Omega_c = 0$) for a controlled duration $\tau_{off}$, while the probe remains continuously applied. During this interval, the atomic state evolves under altered excitation and dissipation conditions before the coupling field is restored. As illustrated conceptually in Fig. 1c, brief interruptions may preserve the preceding oscillation phase, whereas longer interruptions can lead to phase resetting.

To determine whether the recovered oscillation retains its preceding phase, we compare the measured phase after restoration with that expected from uninterrupted evolution. The pre-interruption oscillation is fitted to determine its instantaneous phase and frequency, which are extrapolated across the extinction interval to obtain the predicted phase $\phi_{pred}$. Independently, the phase of the recovered oscillation, $\phi_{rec}$, is extracted from the post-restoration signal and referenced to the same time. Their difference, $\Delta\phi = \phi_{rec} - \phi_{pred}$, quantifies the departure from phase-continuous evolution. Representative measurements in Fig. 1d–g reveal a pronounced dependence on interruption duration. Following a 0.59 μs interruption, the recovered oscillation closely follows the extrapolated trajectory, with $\Delta\phi = -0.2°$. At 3.05 and 5.01 μs, phase displacements of $+19.4°$ and $+38.3°$ emerge, respectively, while a 10.01 μs interruption produces a substantially larger deviation of $-131.7°$. These traces illustrate the progression from nearly phase-continuous recovery to substantial displacement from the preceding oscillatory trajectory. The actual optical extinction intervals are independently measured from the coupling-field signal, with timing jitter of approximately 15–37 ns and close agreement between programmed and measured durations (Supplementary Fig. 1). These measurements establish the experimental basis for distinguishing phase memory from phase resetting as the interruption duration increases.

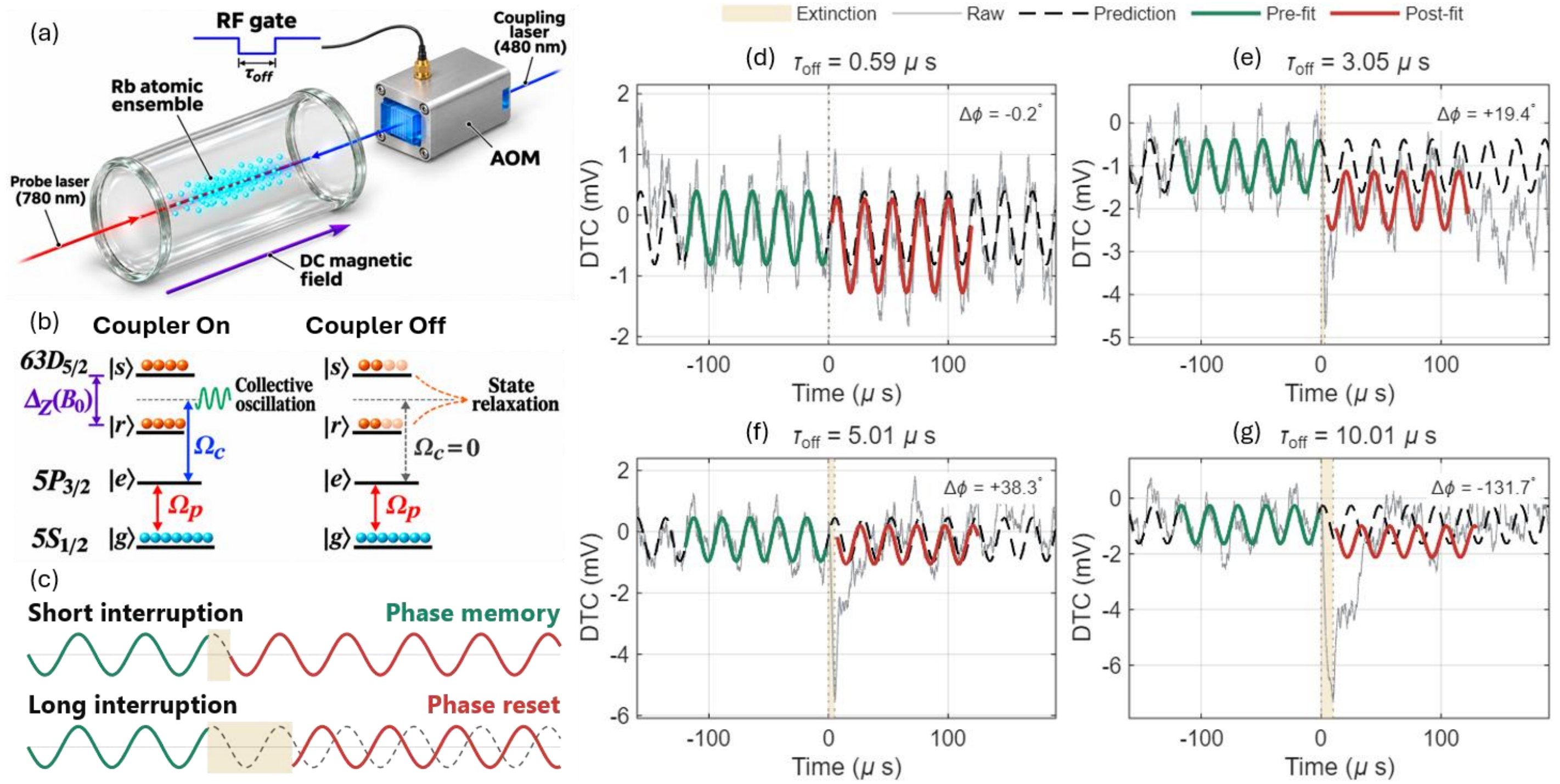

**Fig. 1 | Experimental protocol and observation of phase memory in a dissipative time crystal. a,** Experimental configuration comprising counter-propagating probe (780 nm) and coupling (480 nm) fields addressing a rubidium vapor under a static magnetic field. An acousto-optic modulator (AOM) temporarily extinguishes the coupling field for a duration $\tau_{\text{off}}$. **b,** Energy-level representation under continuous and interrupted excitation. The magnetic field induces Zeeman splitting $\Delta_Z(B_0)$ of the Rydberg manifold, enabling competing excitation pathways and collective oscillations. During interruption ($\Omega_c = 0$), the internal atomic state relaxes while the probe remains on. **c,** Conceptual illustration of phase memory following short interruptions and phase reset following longer interruptions. **d–g,** Representative experimental traces for extinction durations of 0.59, 3.05, 5.01 and 10.01 μs, respectively. Grey curves denote unfiltered measurements, green and red curves denote pre- and post-interruption sinusoidal fits, and black dashed curves represent extrapolated uninterrupted evolution. Shaded regions indicate the measured extinction intervals. The phase deviation $\Delta\phi$ quantifies departure from uninterrupted evolution, illustrating the crossover from phase memory to phase reset.

## Quantifying the lifetime of temporal phase memory

To characterize the persistence of temporal phase information statistically, the interruption protocol is repeated across multiple events while systematically varying $\tau_{\text{off}}$. The event-resolved phase deviations in Fig. 2a remain narrowly distributed around zero for short interruptions but progressively broaden as the extinction duration increases. This behavior is quantified by the phase-continuation fidelity, $F_\phi = \langle \cos\Delta\phi \rangle$, where averaging is performed over repeated interruption events. A value approaching unity indicates faithful phase recovery, whereas values near zero indicate a loss of net phase-continuation fidelity; negative values indicate a net anticorrelation between recovered and predicted phases, without requiring all phase deviations to exceed 90°. As shown in Fig. 2b, $F_\phi$ decreases sharply over the first several microseconds, crossing the $1/e$ threshold at a characteristic phase-memory time of $\tau_{\phi,1/e} = 4.31\,\mu$s. Although the measured fidelity exhibits nonmonotonic variations at intermediate durations, its overall decline establishes the finite temporal persistence of the preceding oscillation phase.

Several controls confirm that this loss of fidelity is not explained by natural phase drift or the phase-extraction procedure. Virtual interruptions introduced into continuously oscillating records retain near-unity fidelity over the corresponding durations (Fig. 2b), with an approximately 8–12° root-mean-square phase error that is largely independent of the virtual-gap duration (Supplementary Fig. 2). Comparable fidelity trends are obtained from unfiltered and independently low-pass-filtered measurements, while variations in post-restoration guard time and fitting-window length preserve the characteristic crossover (Supplementary Fig. 3). These observations establish that the reduction in phase fidelity is associated with interruption of the sustaining excitation rather than accumulated phase uncertainty during uninterrupted evolution or the choice of analysis parameters.

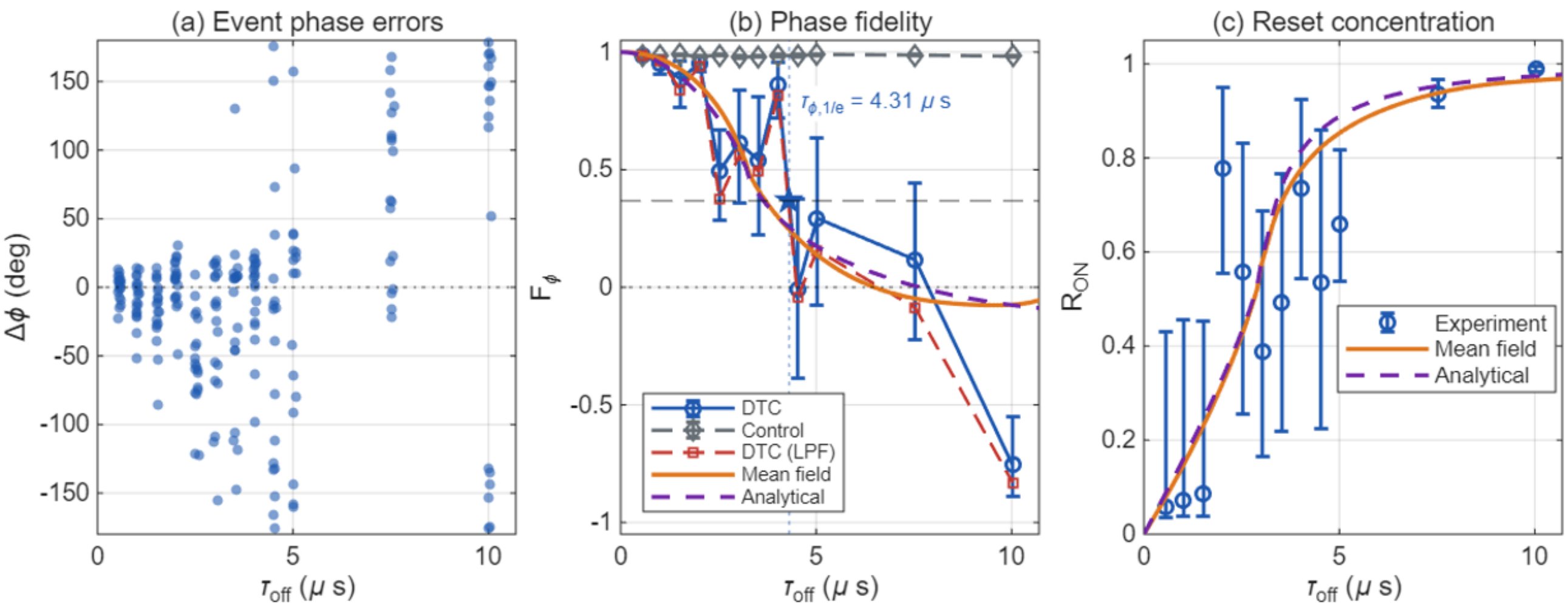

**Fig. 2 | Crossover from phase memory to phase reset. a,** Event-resolved phase deviations $\Delta\phi$ relative to the extrapolated uninterrupted oscillation as a function of coupling-field extinction duration $\tau_{\text{off}}$. The distribution broadens with interruption duration. **b,** Phase fidelity $F_\phi$, defined as the ensemble average of the cosine of the phase deviation. Blue circles represent unfiltered measurements, red squares denote low-pass-filtered (LPF) measurements, and grey diamonds show the control obtained from virtual interruptions of continuous oscillations. The horizontal dashed line marks the $1/e$ fidelity level, yielding a characteristic memory time $\tau_{\phi,1/e} = 4.31$ µs. Negative fidelity indicates preferential recovery with a phase displaced from the extrapolated oscillation. **c,** Reset concentration $R_{\text{ON}}$, quantifying the circular concentration of recovered phases relative to coupling-field restoration. Its increase towards unity indicates increasingly reproducible phase selection upon restart. In **b,c**, solid orange and dashed purple curves show mean-field and analytical predictions, respectively. Together, the two metrics distinguish retention of the pre-interruption phase from deterministic phase selection upon restoration.

Importantly, loss of phase continuation does not necessarily imply random phase recovery. To distinguish phase-memory loss from reproducible phase selection, the circular concentration is evaluated of recovered phases relative to the coupling-field restoration event, $R_{\text{ON}} = |\langle e^{i\phi_{\text{rec}}}\rangle|$, as shown in Fig. 2c. For brief interruptions, the recovered phase closely follows the preceding oscillatory trajectory, yielding a broad distribution of restoration-referenced phases and relatively low concentration. As $\tau_{\text{off}}$ increases, $R_{\text{ON}}$ rises towards unity, indicating increasingly reproducible recovery at a preferred phase relative to the restoration event. The simultaneous reduction in $F_\phi$ and increase in $R_{\text{ON}}$ indicate that the loss of phase continuation is accompanied by increasingly reproducible phase selection rather than random recovery. This behavior is consistent with a crossover from phase memory to phase resetting.

## Phase-transfer maps reveal temporal phase retention

The relationship between the preceding and recovered oscillation phases can be examined more directly through experimental phase-transfer maps, which plot $\phi^*_{\text{rec}}$ against the phase $\phi^*_{\text{pred}}$ predicted from uninterrupted evolution, both expressed modulo 360° (Fig. 3a–f). For extinction durations of 0.56, 1.02 and 2.02 µs, the measurements closely follow the diagonal relation $\phi^*_{\text{rec}} = \phi^*_{\text{pred}}$, indicating that the recovered oscillation preserves its dependence on the phase established before interruption. This near-identity mapping provides direct evidence of temporal phase memory over a range of initial phases, rather than recovery to a single preferred phase. At 2.53 and 5.03 µs, the transfer relation becomes increasingly distorted, with recovered phases departing substantially from the uninterrupted trajectory. By 10.02 µs, the recovered phases concentrate within a comparatively narrow angular range despite variations in the predicted phase. The progression from an approximately diagonal phase-transfer relation towards phase clustering reveals the gradual replacement of history-dependent phase recovery by phase selection associated with renewed excitation.

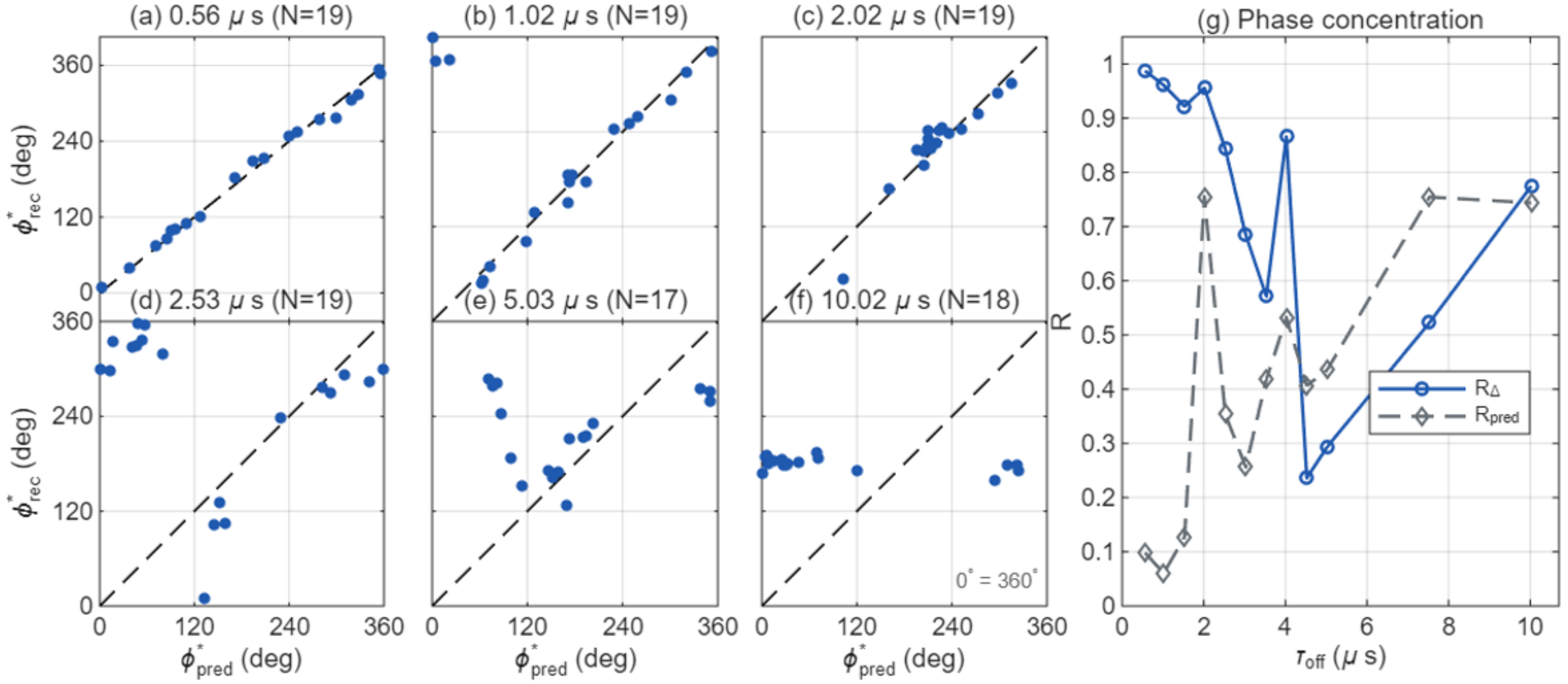

**Fig. 3 | Experimental phase-transfer maps and phase concentration. a–f,** Recovered oscillation phase $\phi^*_{\mathrm{rec}}$ versus the phase predicted by uninterrupted evolution $\phi^*_{\mathrm{pred}}$ for coupling-field extinction durations of 0.56, 1.02, 2.02, 2.53, 5.03 and 10.02 μs, respectively. Each point represents an individual interruption event, with the number of events ($N$) indicated above each panel. Phases are expressed modulo 360°. Black dashed diagonals denote ideal phase continuation. Short interruptions preserve the phase dependence, whereas longer interruptions progressively distort the phase-transfer relation, ultimately producing recovered phases concentrated near a common value. **g,** Circular resultant lengths of the phase deviations ($R_\Delta$, blue circles) and predicted phases ($R_{\mathrm{pred}}$, grey diamonds) as a function of extinction duration $\tau_{\mathrm{off}}$. The latter quantifies the nonuniformity of the sampled predicted phases. High phase-deviation concentration at short interruptions indicates reproducible phase continuation; its resurgence at long interruptions must be interpreted considering the nonuniform predicted-phase distribution and the phase-reset behavior in **f**.

The phase-transfer distributions are further characterized through the circular resultant lengths of the phase deviations, $R_\Delta = \left|\langle e^{i(\phi_{\mathrm{rec}}-\phi_{\mathrm{pred}})}\rangle\right|$, and predicted phases, $R_{\mathrm{pred}} = \left|\langle e^{i\phi_{\mathrm{pred}}}\rangle\right|$ (Fig. 3g). At short interruptions, high $R_\Delta$ indicates tightly concentrated phase deviations and is consistent with faithful continuation across a broad distribution of predicted phases. At intermediate durations, the reduction in $R_\Delta$ reflects the increasing distortion and dispersion of the phase-transfer relation. Its subsequent increase at longer durations must, however, be considered alongside the nonuniform distribution of predicted phases, characterized by $R_{\mathrm{pred}}$. In particular, concentration of phase deviations alone does not establish phase memory when the sampled predicted phases are themselves clustered. Because the experimental events do not uniformly sample the predicted phase, fidelity at longer interruptions need not approach zero even under reproducible resetting. The phase-transfer maps, together with the complementary fidelity and restoration-phase concentration measurements, therefore distinguish phase retention from deterministic resetting. Collectively, these observations establish a finite temporal memory of the spontaneously selected DTC phase and its progressive replacement by reproducible phase selection upon restoration of the sustaining excitation.

## Nonlinear dynamics and analytical description of phase recovery

To investigate the dynamical origin of phase memory, the nonlinear driven–dissipative V-type mean-field model described in Methods is employed, in which coherent excitation, collective interactions and dissipation support self-sustained oscillations [16,23]. Following the experimental protocol, the effective optical drive is temporarily extinguished and subsequently restored, while the internal dynamical variables evolve continuously (Supplementary Sections S1 and S2). The calculated trajectories reveal a pronounced dependence on interruption duration (Fig. 4). For a brief interruption of 0.5 µs, the oscillation resumes close to its uninterrupted trajectory, with a phase deviation of only +3.1° (Fig. 4d). Increasing the interruption to 5 and 10 µs produces progressively stronger amplitude transients and phase deviations of −14.4° and −51.2°, respectively (Fig. 4e,f). These trajectories demonstrate how the temporary removal of the sustaining drive perturbs the established oscillatory state, with subsequent recovery reflecting the internal dynamical state retained during the interruption.

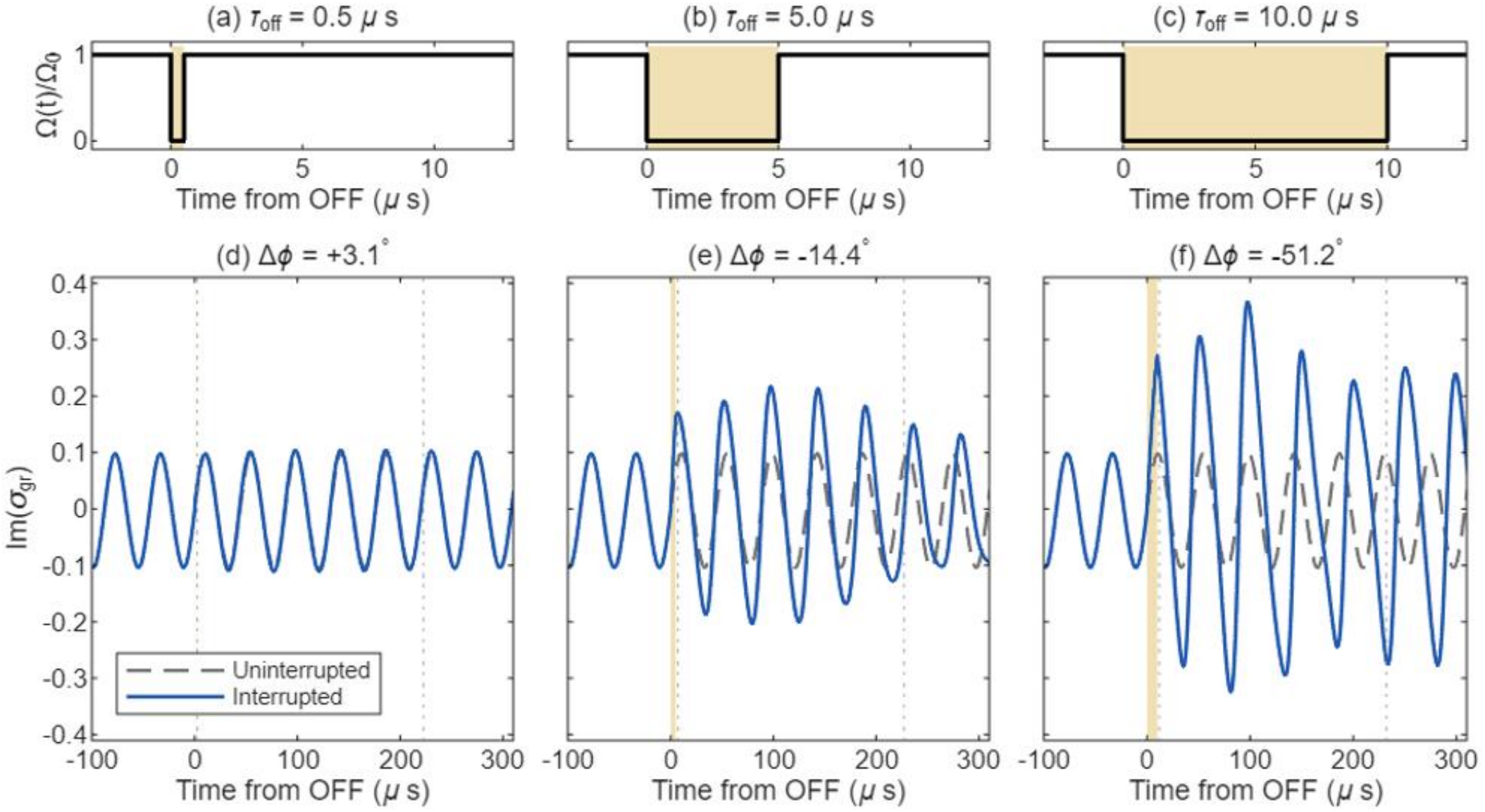

**Fig. 4 | Mean-field dynamics of oscillation recovery following optical interruption. a–c,** Temporal modulation of the effective optical drive, $\Omega(t)/\Omega_0$, for extinction durations of 0.5, 5.0 and 10.0 µs, respectively, where $\Omega_0$ denotes the drive amplitude under continuous excitation. Shaded regions indicate intervals of zero effective drive. **d–f,** Corresponding mean-field trajectories of the imaginary component of the ground–Rydberg coherence, $\mathrm{Im}(\sigma_{gr})$, for interrupted (blue) and uninterrupted (grey dashed) evolution. Time is referenced to the onset of extinction. The recovered phase deviations ($\Delta\phi$) are +3.1°, −14.4° and −51.2°, respectively. Brief interruption produces nearly phase-continuous recovery, whereas longer interruptions generate pronounced amplitude transients and increasing phase displacement. The simulations illustrate how transient evolution of the driven–dissipative atomic state can influence the recovery of collective oscillations.

To determine how this recovery depends on the preceding oscillation phase, we sample 72 uniformly distributed initial positions along the established limit cycle and calculate the phase deviation relative to uninterrupted evolution, as defined in Methods (Fig. 5a; Supplementary Section S2). For short interruptions, the recovered phase remains closely correlated with the original oscillation phase. As the interruption duration increases, the phase response develops pronounced nonlinear distortions, eventually approaching recovery near a preferred phase that depends only weakly on the initial state. Such behavior is consistent with the general framework of phase resetting in nonlinear oscillators [24,25], but here occurs following temporary removal and restoration of the excitation sustaining spontaneous temporal order. The resulting phase-response map reveals a progressive transformation from history-dependent phase recovery to restoration-induced phase selection, rather than simply an increase in random phase uncertainty.

The nonlinear phase-transfer characteristics can be captured by the reduced analytical two-phasor representation introduced in Methods, in which the recovered oscillation is described by the superposition of a decaying retained-memory contribution and a growing restoration-induced contribution. Their relative amplitude, expressed through the memory-to-restart ratio $\eta(\tau_{off})$, provides a compact characterization of the crossover (Fig. 5b; Supplementary Section S3). For $\eta>1$, the retained contribution dominates and the recovered phase remains strongly dependent on its initial value. Near $\eta=1$, the two contributions become comparable, and their near cancellation for particular initial phases produces sharply nonlinear phase responses. As $\eta$ decreases below unity, the restoration-induced contribution becomes dominant and the recovered phase progressively approaches a common value. The effective ratios obtained from individual mean-field phase-transfer maps closely follow the globally fitted analytical dependence, suggesting that the higher-dimensional nonlinear recovery dynamics admit a comparatively simple description through competition between retained phase information and restoration-induced phase selection.

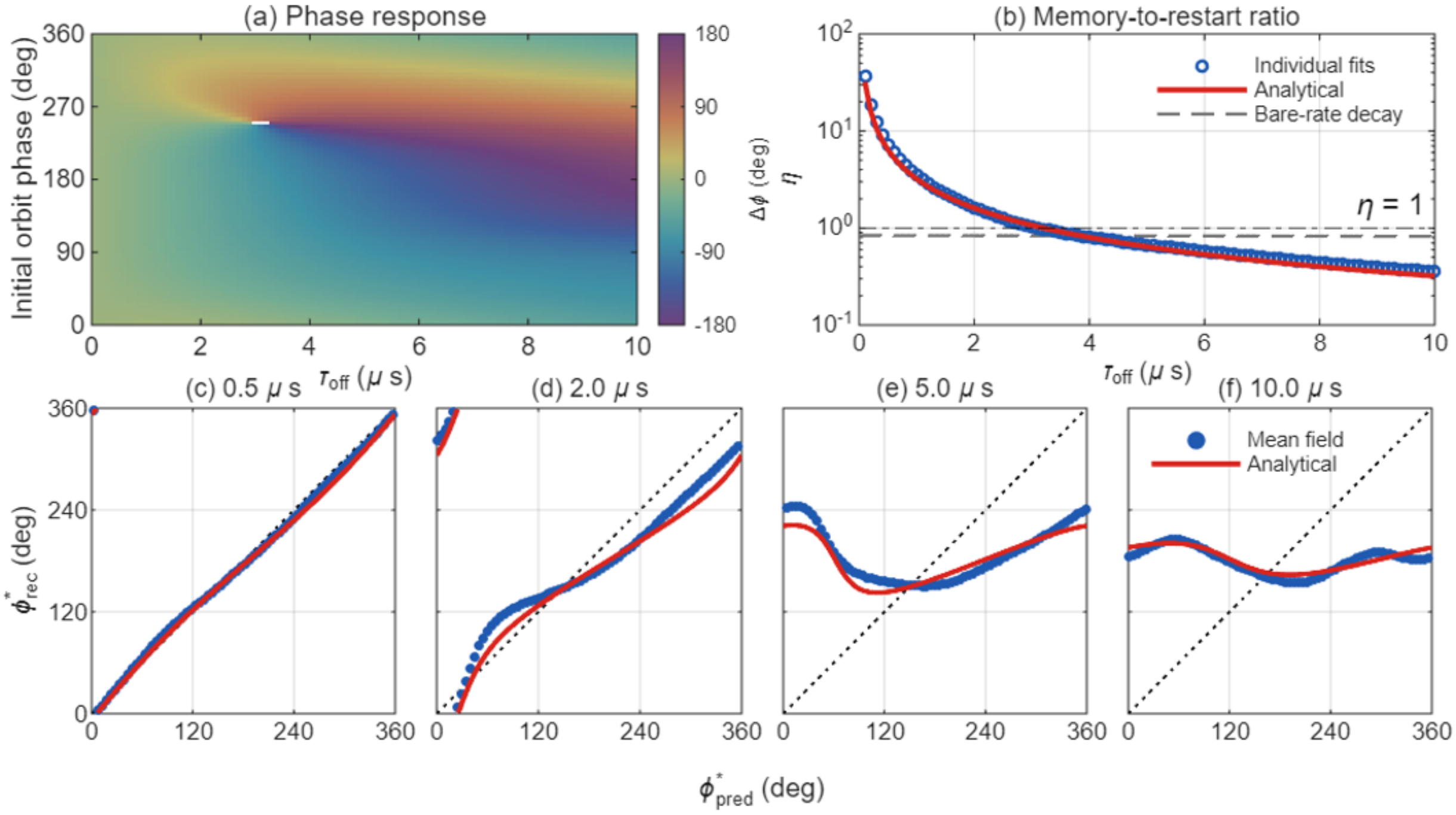

**Fig. 5 | Analytical description of phase memory and restart dynamics. a,** Mean-field phase response $\Delta\phi$ as a function of the initial phase on the oscillatory orbit and coupling-field extinction duration $\tau_{\mathrm{off}}$. The color scale represents the deviation from uninterrupted evolution, revealing a strong dependence on initial phase as the interruption duration increases. **b,** Effective memory-to-restart amplitude ratio $\eta$ extracted from individual fits to the mean-field phase-transfer maps (blue circles), compared with the analytical duration dependence (red curve) and the bare-rate exponential decay reference (grey dashed curve). The horizontal reference $\eta = 1$ marks equal contributions from the retained phase and the restart response. **c–f,** Recovered phase $\phi^*_{\mathrm{rec}}$ versus the predicted uninterrupted phase $\phi^*_{\mathrm{pred}}$ for extinction durations of 0.5, 2.0, 5.0 and 10.0 µs, respectively. Blue points denote mean-field results and red curves show the analytical model; black dotted diagonals indicate ideal phase continuation. The analytical description captures the nonlinear deformation of the phase-transfer maps and their evolution towards a weak dependence on the pre-interruption phase as the restart contribution becomes dominant.

The correspondence between the numerical and analytical descriptions is illustrated by the phase-transfer relations at four interruption durations (Fig. 5c–f). Using a single set of effective parameters calibrated globally against the mean-field phase-transfer maps (Supplementary Section S4), the analytical model reproduces the progression from nearly identity-like phase continuation at 0.5 µs, through strongly nonlinear mappings at intermediate durations, to weak dependence on the initial phase at 10 µs. The same analytical phase-transfer relations generate the phase-continuation fidelity $F_\phi$ and restoration-phase concentration $R_{ON}$ shown alongside the numerical and experimental results in Fig. 2b,c, using the observables defined in Methods. The close agreement between the reduced analytical and mean-field phase-transfer maps suggests that the crossover can be understood through a low-dimensional competition between retained phase information and restoration-induced phase selection, despite the underlying collective nonlinear dynamics. Although the two-phasor representation is calibrated to the mean-field calculations rather than derived directly from the microscopic equations, it provides a unified interpretation of phase preservation, nonlinear phase transfer and deterministic resetting in a dissipative time crystal.

## DISCUSSION

The present findings demonstrate that a dissipative time crystal can retain information about its spontaneously established oscillation phase following temporary interruption of the excitation sustaining its collective dynamics. This observation identifies phase memory as a distinct dynamical property of time-crystalline order, extending its characterization beyond persistent oscillations and frequency stability. The finite memory timescale further shows that retention of temporal phase information can be quantified independently of the subsequent recovery of collective oscillations.

The observed crossover from phase-coherent recovery to reproducible phase resetting distinguishes the loss of temporal information from randomization of the oscillation phase. Experimentally, this distinction emerges through the complementary evolution of phase-continuation fidelity and restoration-phase concentration. Nonlinear mean-field calculations reproduce the principal features of this crossover, while a reduced analytical description captures the phase-transfer dynamics through competition between retained phase information and restoration-induced phase selection. This correspondence suggests that the recovery of collective temporal order can exhibit an effectively low-dimensional structure despite its underlying nonlinear dynamics, although the microscopic pathways governing phase retention remain to be identified.

More broadly, phase memory provides a means of investigating the stability and controllability of nonequilibrium temporal order. Future theoretical studies could examine how interactions, driving and dissipation govern memory timescales and nonlinear phase-transfer dynamics, while more detailed microscopic descriptions may clarify the mechanisms of phase retention. Experimental investigations of controllably coupled time-crystalline ensembles could further reveal whether temporal phase information can be preserved, transferred or reconfigured. These directions build on the experimentally demonstrated capacity of dissipative time-crystalline order to retain a memory of its preceding phase, opening possibilities for phase-coherent control in driven–dissipative many-body systems.

## METHODS

### Experimental apparatus and optical interruption

The experimental system follows the magnetically initiated Rydberg dissipative time-crystal configuration described previously [29], employing the same room-temperature, isotopically enriched $^{87}$Rb vapor cell (56 mm length, 25 mm diameter), optical excitation arrangement and Helmholtz coil assembly. Counterpropagating probe (780 nm) and coupling (approximately 480 nm) fields address the $5S_{1/2}$, F = 2 → $5P_{3/2}$, F′ = 3 → $63D_{5/2}$ ladder. The probe is generated by a tunable diode laser (TOPTICA DL pro) stabilized using Doppler-free saturated-absorption spectroscopy, while the coupling field is produced by second-harmonic generation of a tunable 960 nm laser using a TOPTICA SHG system. The laser linewidths are below 90 kHz. The probe and coupling powers delivered to the vapor cell are approximately 312 μW and 430 mW, respectively, with beam diameters of approximately 0.8 mm ($1/e^2$ intensity). A static magnetic field of approximately 19 G is applied along the optical axis using a Helmholtz coil pair (Eisco Labs, PH0845HEL). Under these experimental conditions, the dissipative time crystal exhibits self-sustained oscillations in probe transmission at a nominal frequency of approximately 42.075 kHz.

Rapid interruption of Rydberg excitation is implemented using a free-space acousto-optic modulator (AeroDIODE 532FSAOM-200-1.0), operated at a radiofrequency of 200 MHz. The coupling beam passes through the approximately 1 mm active aperture, and an iris selects the first-order diffracted beam while rejecting undiffracted light. A high-speed waveform generator supplies rectangular pulses to gate the AOM radiofrequency drive, thereby interrupting the coupling field while the probe remains continuously applied. A fraction of the coupling light is diverted to a 2 GHz bandwidth Thorlabs photodetector to independently monitor the optical switching response. During interruption, the detected optical signal reaches the photodetector background noise, indicating extinction within the measurement sensitivity. Measured interruption durations closely follow their programmed values, with shot-to-shot timing jitter of approximately 15–37 ns (Supplementary Fig. 1).

The transmitted probe light is detected using an 11 MHz bandwidth photodetector to monitor the collective DTC oscillations. Its electrical output is divided into two acquisition paths: one is recorded directly without additional filtering, while the other passes through a 50 kHz low-pass filter before acquisition. Both signals are sampled simultaneously on separate channels of a digital oscilloscope at approximately 1.5 GSa/s, alongside the coupling-field monitor. The oscilloscope acquisition is triggered using the high-speed coupling-field photodetector signal, providing a common timing reference for the optical interruption and the pre- and post-interruption DTC oscillations. The unfiltered probe signal is used for the primary phase-memory analysis, while the filtered channel provides an independent comparison of the extracted phase dynamics (Supplementary Figs. 2 and 3).

**Nonlinear V-type mean-field model**

Interruption-induced phase memory is investigated using the nonlinear driven–dissipative V-type mean-field framework introduced in Refs. [16, 23]. An effective ground state $|g\rangle$ couples to two excited states $|r\rangle$ and $|s\rangle$. The state is specified by excited-state populations $n_r$ and $n_s$, optical coherences $\sigma_{gr}$ and $\sigma_{gs}$, and excited-state coherence $\sigma_{rs}$. With microwave injection absent ($\Omega_{rs} = 0$), the equations implemented in our simulations are

$$
\begin{aligned}
\dot{n}_r &= \Omega(t)/2\,\mathrm{Im}(\sigma_{gr}) - \gamma n_r, \\
\dot{n}_s &= \Omega(t)/2\,\mathrm{Im}(\sigma_{gs}) - \gamma n_s, \\
\dot{\sigma}_{gr} &= i\Omega(t)/2\,(2n_r + n_s + \sigma_{rs} - 1) + \big[i\big(\Delta_r - \chi(n_r + n_s)\big) - \gamma/2\big]\sigma_{gr}, \\
\dot{\sigma}_{gs} &= i\Omega(t)/2\,(2n_s + n_r + \sigma_{rs} - 1) + \big[i\big(\Delta_s - \chi(n_r + n_s)\big) - \gamma/2\big]\sigma_{gs}, \\
\dot{\sigma}_{rs} &= i\Omega(t)/2\,\big(\sigma_{gs} - \sigma_{gr}\big) - [\gamma + i(\Delta_r - \Delta_s)]\sigma_{rs}.
\end{aligned}
$$

Here $\Omega(t)$ is the effective optical drive, $\gamma$ is the dissipation rate, $\Delta_r$ and $\Delta_s$ are branch detunings, and $\chi$ sets the population-dependent nonlinear detuning $\chi(n_r + n_s)$. We define $\delta \equiv \Delta_r - \Delta_s$ and $\chi_0 \equiv -16\delta$, the nominal interaction coefficient of the reference model; the present calculations use $\chi = 0.7\chi_0$. Thus the coefficient is reduced by 30% in magnitude relative to that baseline, an experimentally informed adjustment rather than a parameter-free prediction. The simulated oscillatory observable is $\mathrm{Im}(\sigma_{gr})$. This effective V-type framework captures DTC behavior [16, 23, 30-31] and collective nonlinear oscillations but does not explicitly resolve the separate probe and coupler fields of the experimental ladder system.

**Interruption protocol and phase observables**

After convergence to a stable limit cycle, the drive is set to $\Omega(t) = 0$ for an interval $\tau_{\mathrm{off}}$ and then restored to $\Omega_0$. All five state variables evolve continuously during extinction and recovery; none is reset selectively. We sample 72 equally spaced initial positions along the settled orbit and propagate an uninterrupted reference trajectory from each initial state. As in the experiment, five-cycle sinusoidal fits extract the fundamental phase before and after interruption, referenced to the optical restoration time. The phase error and two summary measures are

$$
\begin{aligned}
\Delta\phi_j &= \mathrm{Arg}\Big[e^{i(\phi_{\mathrm{rec},j} - \phi_{\mathrm{pred},j})}\Big], \\
F_\phi(\tau) &= \frac{1}{N}\sum_{j=1}^{N} \cos\Delta\phi_j, \qquad R_{\mathrm{ON}}(\tau) = \left|\frac{1}{N}\sum_{j=1}^{N} e^{i\phi_{\mathrm{rec},j}}\right|.
\end{aligned}
$$

Here $\phi_{\mathrm{pred},j}$ is the phase predicted by uninterrupted continuation, whereas $\phi_{\mathrm{rec},j}$ is the recovered phase. $F_\phi$ tests phase continuity and $R_{\mathrm{ON}}$ tests concentration around a common restoration phase; they need not change in the same way.

**Reduced analytical phase-recovery model**

To interpret the nonlinear phase-transfer maps, we approximate recovery by the sum of a retained-memory phasor and a growing restoration-induced phasor,

$$
\begin{aligned}
Z_{\mathrm{rec}}(\phi_{\mathrm{pred}},\tau) &= \eta_0 e^{-\Gamma\tau} e^{i[\phi_{\mathrm{pred}}+(\omega_m-\omega_0)\tau]} + (1-e^{-\Gamma\tau})e^{i\phi_r},\\
\phi_{\mathrm{rec}}^{\mathrm{AN}} &= \mathrm{Arg}(Z_{\mathrm{rec}}), \qquad \eta(\tau) = \frac{\eta_0 e^{-\Gamma\tau}}{1-e^{-\Gamma\tau}}.
\end{aligned}
$$

The parameters $\eta_0$ and $\Gamma$ describe the effective retained amplitude and its relaxation, $\omega_m - \omega_0$ represents phase evolution relative to the free-running DTC, and $\phi_r$ is the preferred restoration phase. The ratio $\eta \gg 1$ implies memory-dominated recovery, $\eta \sim 1$ strongly nonlinear phase transfer, and $\eta \ll 1$ reset-dominated recovery. The analytical phase in the equation above is used to calculate the curves in Figs. 2 and 5 through the same definitions of $F_\phi$ and $R_{\mathrm{ON}}$. A single global circular-error fit of $\eta_0$, $\Gamma$, and $\omega_m$ to the mean-field, not experimental, phase-transfer maps sets these curves; $\phi_r$ is estimated from the long-interruption simulated response. Numerical parameters, fitting details, and the bare-rate comparison appear in the Supplemental Material. The two-phasor relation is an explanatory approximation, not an independent microscopic derivation.

## REFERENCES

1. Wilczek, F. Quantum time crystals. *Phys. Rev. Lett.* **109**, 160401 (2012).
2. Shapere, A. & Wilczek, F. Classical time crystals. *Phys. Rev. Lett.* **109**, 160402 (2012).
3. Watanabe, H. & Oshikawa, M. Absence of quantum time crystals. *Phys. Rev. Lett.* **114**, 251603 (2015).
4. Else, D. V., Bauer, B. & Nayak, C. Floquet time crystals. *Phys. Rev. Lett.* **117**, 090402 (2016).
5. Yao, N. Y., Potter, A. C., Potirniche, I.-D. & Vishwanath, A. Discrete time crystals: rigidity, criticality, and realizations. *Phys. Rev. Lett.* **118**, 030401 (2017).
6. Zhang, J. et al. Observation of a discrete time crystal. *Nature* **543**, 217–220 (2017).
7. Choi, S. et al. Observation of discrete time-crystalline order in a disordered dipolar many-body system. *Nature* **543**, 221–225 (2017).
8. Mi, X. et al. Time-crystalline eigenstate order on a quantum processor. *Nature* **601**, 531–536 (2022).
9. Diehl, S. et al. Quantum states and phases in driven open quantum systems with cold atoms. *Nat. Phys.* **4**, 878–883 (2008).
10. Iemini, F. et al. Boundary time crystals. *Phys. Rev. Lett.* **121**, 035301 (2018).
11. Buča, B., Tindall, J. & Jaksch, D. Non-stationary coherent quantum many-body dynamics through dissipation. *Nat. Commun.* **10**, 1730 (2019).
12. Keßler, H. et al. Observation of a dissipative time crystal. *Phys. Rev. Lett.* **127**, 043602 (2021).
13. Kongkhambut, P. et al. Observation of a continuous time crystal. *Science* **377**, 670–673 (2022).
14. Chen, Y.-H. & Zhang, X. Realization of an inherent time crystal in a dissipative many-body system. *Nat. Commun.* **14**, 6161 (2023).
15. Greilich, A. et al. Robust continuous time crystal in an electron–nuclear spin system. *Nat. Phys.* **20**, 631–636 (2024).
16. Wu, X. et al. Dissipative time crystal in a strongly interacting Rydberg gas. *Nat. Phys.* **20**, 1389–1394 (2024).
17. Adler, R. A study of locking phenomena in oscillators. *Proc. IRE* **34**, 351–357 (1946).
18. Walter, S., Nunnenkamp, A. & Bruder, C. Quantum synchronization of a driven self-sustained oscillator. *Phys. Rev. Lett.* **112**, 094102 (2014).
19. Wadenpfuhl, K. & Adams, C. S. Emergence of synchronization in a driven-dissipative hot Rydberg vapor. *Phys. Rev. Lett.* **131**, 143002 (2023).
20. Kongkhambut, P. et al. Observation of a phase transition from a continuous to a discrete time crystal. *Rep. Prog. Phys.* **87**, 080502 (2024).
21. Liu, B. et al. Bifurcation of time crystals in driven and dissipative Rydberg atomic gas. *Nat. Commun.* **16**, 1419 (2025).
22. Greilich, A. et al. Exploring nonlinear dynamics in periodically driven time crystal from synchronization to chaotic motion. *Nat. Commun.* **16**, 2936 (2025).
23. Arumugam, D. Injection locking of Rydberg dissipative time crystals. *Commun. Phys.* **9**, 156 (2026).
24. Glass, L. & Winfree, A. T. Discontinuities in phase-resetting experiments. *Am. J. Physiol.* **246**, R251–R258 (1984).

25. Krishnan, G. P., Bazhenov, M. & Pikovsky, A. Multipulse phase resetting curves. *Phys. Rev. E* **88**, 042902 (2013).
26. Carr, C., Ritter, R., Wade, C., Adams, C. S. & Weatherill, K. J. Nonequilibrium phase transition in a dilute Rydberg ensemble. *Phys. Rev. Lett.* **111**, 113901 (2013).
27. Marcuzzi, M., Levi, E., Diehl, S., Garrahan, J. P. & Lesanovsky, I. Universal nonequilibrium properties of dissipative Rydberg gases. *Phys. Rev. Lett.* **113**, 210401 (2014).
28. Ding, D. et al. Ergodicity breaking from Rydberg clusters in a driven-dissipative many-body system. *Sci. Adv.* **10**, eadl5893 (2024).
29. Arumugam, D. Microwave power-to-frequency transduction in a magnetically initiated Rydberg dissipative time crystal. Preprint at https://arxiv.org/abs/2609.31824
30. Arumugam, D. Electric-field sensing with driven-dissipative time crystals in room-temperature Rydberg vapor. *Sci. Rep.* **15**, 13446 (2025).
31. Arumugam, D. Stark modulated Rydberg dissipative time crystals at room temperature applied to sub-kHz electric field sensing. *Sci. Rep.* **15**, 35976 (2025).

**Acknowledgments**

The research was carried out at the Jet Propulsion Laboratory, California Institute of Technology, under a contract with the National Aeronautics and Space Administration (80NM0018D0004), through the Instrument Incubator Program's (IIP) Instrument Concept Development (80NM0022F0020).

**Funding:**

National Aeronautics and Space Administration grant 80NM0018D0004
National Aeronautics and Space Administration grant 80NM0022F0020

**Author contributions:** D.A conceived of the study and experiments, configured the atomic systems to include lasers and locking systems, and collected and processed all data reported in the text and figures. D.A. also developed all modeling, theoretical derivations and numerical simulations used or reported.

**Competing interests:** The author declares no competing interest.

**Data, code, and materials availability:** The data and computational codes supporting the findings of this study are available upon reasonable request during peer review and will be deposited in a public repository prior to publication.

## Supplementary Materials
## Part I — Theoretical and numerical methods

### S1. Effective model and physical parameters

The main text gives the five coupled equations for the phenomenological V-type model, with state vector $\mathbf{y} = (n_r, n_s, \sigma_{gr}, \sigma_{gs}, \sigma_{rs})^{\mathrm{T}}$. The two excited-state populations produce the common nonlinear detuning $\chi(n_r + n_s)$. Microwave coupling is absent ($\Omega_{rs} = 0$). The effective drive $\Omega(t)$ represents optical excitation collectively; setting it to zero approximates coupler extinction in the experiment, although the probe remains on. Thus, these equations are not a full microscopic model of the ladder-level experiment or a positivity-preserving Lindblad evolution.

The numerical parameters (angular-frequency units, $\mathrm{s}^{-1}$) are:

$\gamma = 25.4 \times 10^3/(2\pi)$; $\Omega_0 = (0.85 \times 10^6)(13.97 \times 10^6)/(2 \times 22 \times 10^6 \times 2\pi)$; $\Delta_r = -0.34 \times 10^6/(2\pi)$; $\delta = 1.069 \times 10^6/(2\pi)$; $\Delta_s = \Delta_r - \delta$; and $\chi = 0.7\chi_0$, where $\chi_0 = -16\delta$.

Here $\chi_0$ is the nominal interaction coefficient in the reference mean-field parameterization, not an independently measured interaction strength. The 0.7 scaling is informed by the experimental phase-memory comparison and should not be interpreted as a parameter-free prediction. The model structure follows Refs. [16, 23]; all quantities necessary to run the present simulation are specified here and in the main text.

## S2. Interruption simulation and phase reconstruction

The coupled nonlinear differential equations were integrated using an adaptive-step Dormand–Prince Runge–Kutta (4,5) method, with relative and absolute error tolerances of $10^{-8}$ and $10^{-10}$, respectively, and a maximum integration step of $0.5\,\mu s$. A 30 ms continuously driven trajectory is used to reach an oscillatory state. The free-running period $T_0 = 1/f_0$ is estimated from upward zero crossings of the mean-subtracted signal $S(t) = \mathrm{Im}\sigma_{gr}(t)$ within its final 5 ms. Seventy-two equally spaced times on the final orbit define the initial-phase ensemble. Settling and periodicity are checked before proceeding.

For each initial state, the model first evolves without optical drive for $\tau_{\mathrm{off}}$, then with $\Omega_0$ restored, without resetting any of the five state variables. An uninterrupted reference is propagated from the identical starting state. The OFF-duration grid is 0–12 $\mu s$ in $0.1\,\mu s$ increments; the published phase-response map displays its 0–10 $\mu s$ portion. Post-restoration evolution spans $500\,\mu s$.

The pre- and post-interruption fundamental phases are estimated independently by linear least squares using

$$S_{\mathrm{fit}}(t) = c_0 + a\cos[2\pi f_0(t - t_{\mathrm{ref}})] + b\sin[2\pi f_0(t - t_{\mathrm{ref}})], \qquad \phi = \mathrm{atan2}(-b, a).$$

Each fit spans five periods, with $2\,\mu s$ pre-interruption and post-restoration guards in the primary analysis. Fits with $R^2 < 0.05$ are rejected. Additional post-restoration guards of 10, 25, 50, and 75 $\mu s$ test sensitivity to the recovery window. Extrapolating the pre-interruption fit to restoration gives $\phi_{\mathrm{pred}}$; fitting the recovered signal gives $\phi_{\mathrm{rec}}$. Both are referred to the same time origin, and the wrapped error is

$$\Delta\phi = \mathrm{Arg}\left[e^{i(\phi_{\mathrm{rec}} - \phi_{\mathrm{pred}})}\right].$$

The main-text observables are $F_\phi = \langle \cos\Delta\phi \rangle$ and $R_{\mathrm{ON}} = \left|\langle e^{i\phi_{\mathrm{rec}}} \rangle\right|$. In particular, $F_\phi$ is a mean cosine, not a phase-acceptance fraction. A negative value indicates anticorrelation with predicted phase.

## S3. Reduced analytical phase transfer

The reduced ansatz introduced in the main text superposes a retained-memory phasor and a restoration-induced phasor. For nonzero interruption duration, its recovery phase may equivalently be written as

$$\phi_{\mathrm{rec}}^{\mathrm{AN}} = \phi_r + \mathrm{Arg}\left[1 + \eta(\tau)e^{i\theta}\right], \qquad \theta = \phi_{\mathrm{pred}} + (\omega_m - \omega_0)\tau - \phi_r,$$

where $\eta(\tau) = \eta_0 e^{-\Gamma\tau}/(1 - e^{-\Gamma\tau})$ and $\omega_0 = 2\pi f_0$. Equivalently,

$$\phi_{\mathrm{rec}}^{\mathrm{AN}} = \phi_r + \mathrm{atan2}(\eta\sin\theta, 1 + \eta\cos\theta).$$

The growing restoration term enforces $\phi_{\mathrm{rec}}^{\mathrm{AN}} \to \phi_{\mathrm{pred}}$ as $\tau \to 0$. When $\eta \ll 1$, $\phi_{\mathrm{rec}}^{\mathrm{AN}} \simeq \phi_r + \eta\sin\theta$: the reset-dominated response retains a small phase-dependent modulation. Near $\eta = 1$, the phasors have comparable magnitudes, giving strongly distorted phase-transfer maps. This construction is an interpretable approximation to the simulated map, not an analytical solution of the five nonlinear differential equations.

## S4. Calibration and restricted comparisons

The preferred reset phase $\phi_r$ is the circular mean of the simulated recovery phases at the longest interruption ($12\,\mu s$). Holding $\phi_r$ fixed, a global fit of $\eta_0$, the positive effective rate $\Gamma$, and the effective OFF-state rotation $\omega_m$ uses all valid mean-field phase pairs with $0 < \tau \le 10\,\mu s$ and minimizes

$$\mathcal{L} = \left\langle 1 - \cos(\phi_{\mathrm{rec}}^{\mathrm{MF}} - \phi_{\mathrm{rec}}^{\mathrm{AN}}) \right\rangle.$$

The effective parameters were optimized using a multistart Nelder–Mead simplex method, with parameter bounds enforced through a penalized objective function. It also fits a separate $\eta$ at each duration, keeping $\omega_m$ and $\phi_r$ fixed; these values are diagnostics of the global reduction (Fig. 5b), not independent physical measurements.

Two restricted alternatives assess the importance of effective rates: (i) a constant restoration phasor with exponentially decaying retained amplitude and (ii) a growing restoration phasor constrained to the bare OFF-state optical-coherence damping $\gamma/2$ and linear rotation $\Delta_r$. The fully effective model releases the relaxation and rotation rates. Because the OFF-state dynamics retain a population-dependent shift, $\gamma/2$ and $\Delta_r$ cannot in general be identified with the emergent oscillator's phase-memory decay and rotation rates.

The analytical curves in Fig. 2 are calculated from $F_\phi^{\mathrm{AN}} = \langle\cos(\phi_{\mathrm{rec}}^{\mathrm{AN}} - \phi_{\mathrm{pred}})\rangle$ and $R_{\mathrm{ON}}^{\mathrm{AN}} = \left|\langle e^{i\phi_{\mathrm{rec}}^{\mathrm{AN}}}\rangle\right|$ over the uniform simulated initial-phase ensemble. The same global analytical parameters generate the cuts in Fig. 5. These comparisons assess the descriptive accuracy of the reduced model relative to the calibrated mean-field dynamics; they do not constitute an independent parameter-free prediction of experiment.

## Part II. Experimental characterization and validation

### S5. Optical extinction and interruption timing

The temporal characteristics of the coupling-field interruption were independently measured using a 2 GHz bandwidth photodetector monitoring a fraction of the optical power delivered through the acousto-optic modulator (AOM). The photodetector signal was acquired simultaneously with the AOM drive waveform, allowing the actual optical extinction interval to be distinguished from the programmed electrical gating duration. During interruption, the detected coupling-field signal decreased to the photodetector background noise, confirming optical extinction within the measurement sensitivity.

The optical interruption duration, $\tau_{off}$, was determined from the falling and rising crossings of a fixed 2 V threshold applied to the coupling-field monitor signal (Supplementary Fig. 1a). For the representative event, the measured extinction interval was 1.52 µs. The shot-to-shot timing jitter, quantified by the standard deviation of the measured interruption duration at each programmed setting, ranged from approximately 15 to 37 ns across the investigated durations (Supplementary Fig. 1b). Furthermore, the measured optical extinction intervals closely followed the programmed gate durations, $\tau_{set}$, over the approximately 0.5–10 µs range (Supplementary Fig. 1c). These measurements confirm that the optical interruptions are reproducible and accurately characterized on timescales substantially shorter than the observed phase-memory crossover. The measured optical transitions, rather than the nominal electrical gating waveform, were therefore used to define the interruption intervals in the phase-memory analysis.

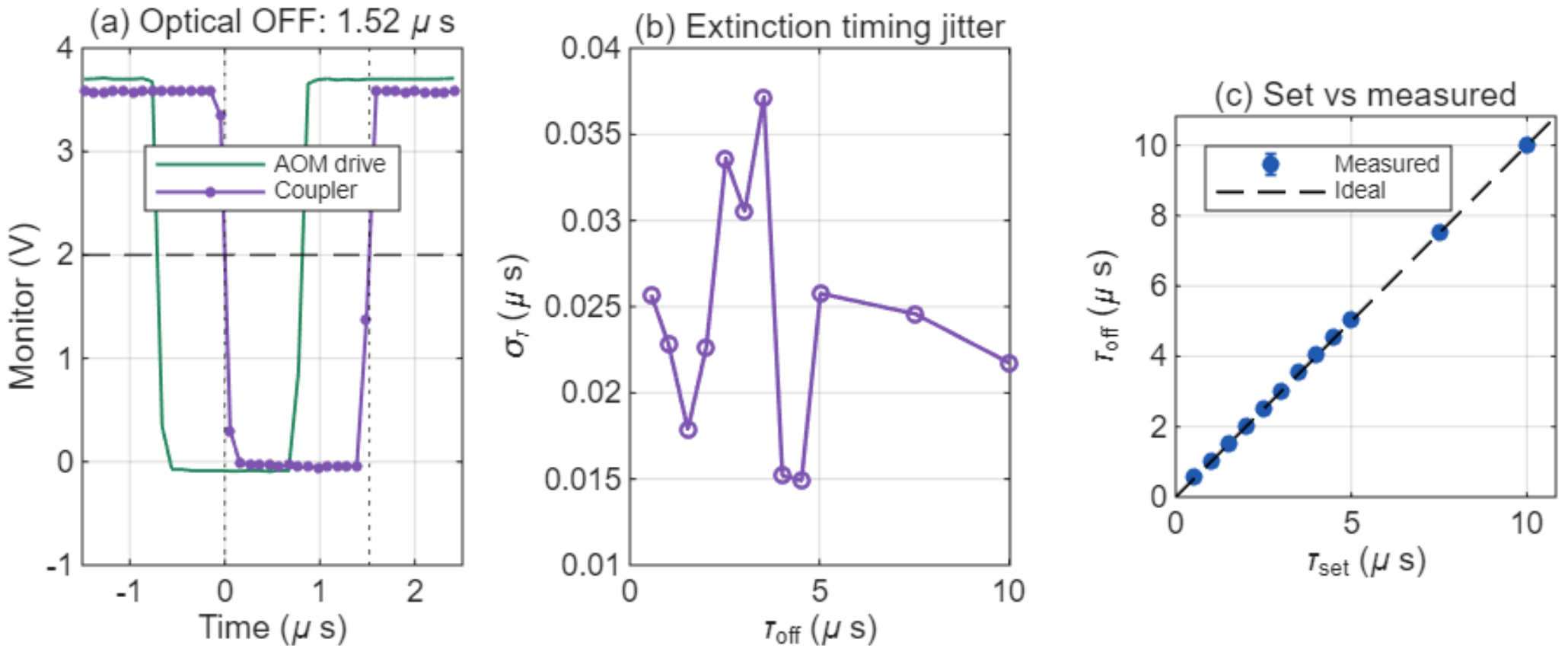

**Supplementary Fig. 1 | Calibration and timing precision of coupling-field extinction. a,** Representative AOM drive (green) and independently monitored coupling-field signal (purple) for an optical extinction interval of 1.52 µs. The extinction duration $\tau_{\mathrm{off}}$ is determined from the falling and rising crossings of the 2 V threshold (horizontal dashed line), indicated by vertical dashed lines. **b,** Shot-to-shot timing jitter $\sigma_\tau$, quantified by the standard deviation of the measured extinction duration, ranges from approximately 15 to 37 ns. **c,** Measured optical extinction duration $\tau_{\mathrm{off}}$ as a function of the programmed gate duration $\tau_{\mathrm{set}}$. The black dashed line denotes ideal equality. The close agreement and low timing jitter demonstrate the temporal precision of the interruption protocol.

## S6. Experimental phase extraction and detection controls

The robustness of the experimental phase measurements was evaluated using simultaneous recordings of the unfiltered and low-pass-filtered DTC signals. As described in Methods, the output of the 11 MHz probe-transmission photodetector was divided into two acquisition channels, one recorded directly and the other filtered using a 50 kHz low-pass filter (LPF). The fundamental oscillation phase was extracted independently from each channel using sinusoidal fits to the pre-interruption and post-restoration signals. Representative waveforms and fitted oscillations are shown in Supplementary Fig. 2a,b. The LPF suppresses higher-frequency fluctuations in the detected signal while preserving the fundamental DTC oscillation, allowing the influence of detection bandwidth on phase reconstruction to be assessed.

The difference between the phase-error estimates obtained from the two detection channels, $\delta\phi_{fit}$, is shown as a function of optical interruption duration in Supplementary Fig. 2c. Most differences remain within approximately ±40° across the investigated range, with substantially smaller deviations for many events. Although the channels exhibit event-dependent differences, the agreement in their overall phase-memory behavior indicates that the observed crossover is not determined by the additional low-pass filtering. The unfiltered measurements are therefore used for the primary analysis, avoiding possible distortion of the transient response by the LPF.

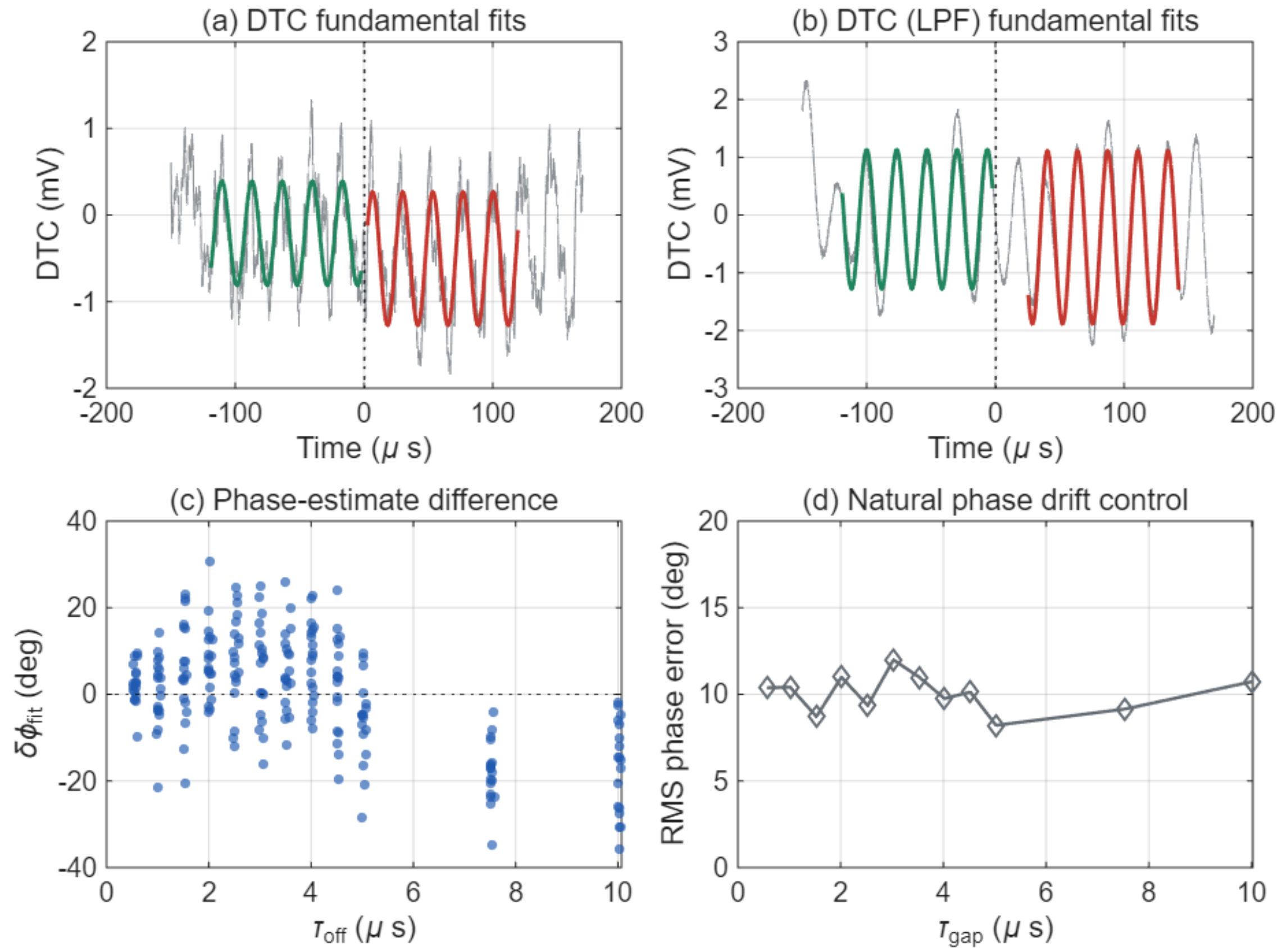

**Supplementary Fig. 2 | Robustness of phase extraction and natural phase drift. a,b,** Representative DTC signals acquired without additional filtering and with a 50 kHz low-pass filter (LPF), respectively. Grey curves denote measured signals, and green and red curves denote sinusoidal fits to the fundamental oscillation before and after interruption. **c,** Event-resolved difference between the phase-error estimates obtained from the two detection channels, $\delta\phi_{\text{fit}}$, versus optical extinction duration $\tau_{\text{off}}$. Differences remain within approximately ±40° across the measured range. **d,** Root-mean-square phase error extracted from uninterrupted DTC oscillations using virtual gaps of duration $\tau_{\text{gap}}$. The approximately 8–12° baseline phase error shows little dependence on gap duration, indicating that natural phase drift alone cannot account for the loss of phase memory observed under optical interruption.

To distinguish interruption-induced phase-memory loss from intrinsic oscillator phase drift or errors in phase reconstruction, a separate control analysis was performed using uninterrupted DTC recordings. Virtual interruption intervals of duration $\tau_{gap}$ were introduced into these records without physically extinguishing the coupling field. The phase before each virtual gap was extrapolated to the corresponding endpoint and compared with the independently fitted subsequent oscillation phase, using the same reconstruction procedure as for actual interruptions. The resulting root-mean-square phase error remains approximately 8–12° over the investigated gap durations, with no systematic increase up to 10 µs (Supplementary Fig. 2d). This comparatively stable baseline demonstrates that the pronounced loss of phase-continuation fidelity observed during actual optical interruptions cannot be attributed to natural phase drift over the interruption timescale or to the phase-extraction procedure alone.

## S7. Robustness of phase-memory analysis

The sensitivity of the phase-memory measurements to the phase-reconstruction procedure was evaluated by varying the post-restoration fitting interval and the number of oscillation cycles used for phase estimation. In the primary analysis, the phase-continuation fidelity $F_\phi$ and restoration-phase concentration $R_{ON}$ were calculated using sinusoidal fits to the fundamental DTC oscillation. The fitting procedure was repeated with post-restoration guard times of 2, 10, 25, 50 and 75 µs, excluding the corresponding interval immediately following optical restoration to assess the influence of transient recovery dynamics.

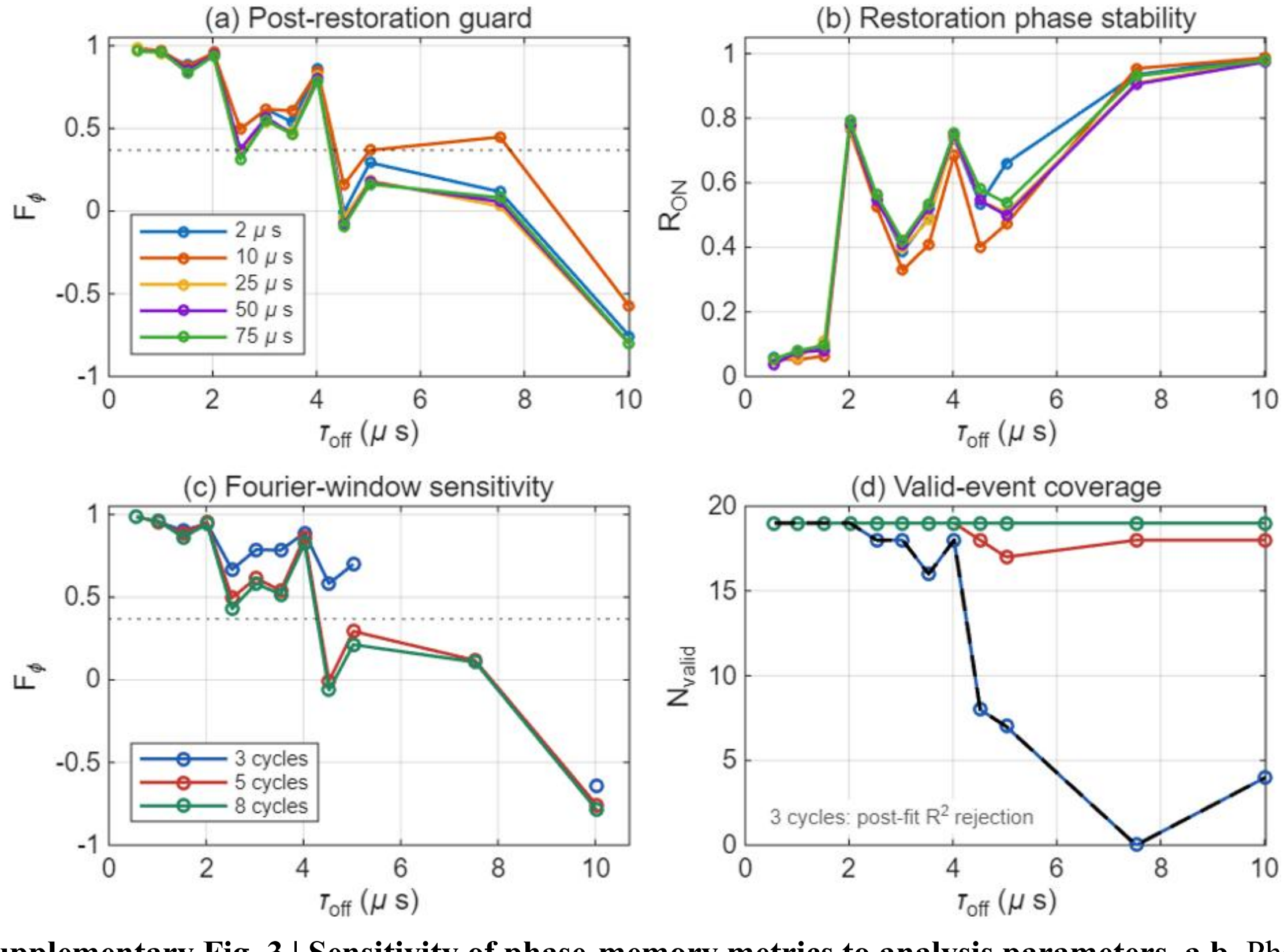

**Supplementary Fig. 3 | Sensitivity of phase-memory metrics to analysis parameters. a,b,** Phase fidelity $F_\phi$ and restoration-phase concentration $R_{\mathrm{ON}}$, respectively, as functions of optical extinction duration $\tau_{\mathrm{off}}$ for post-restoration guard times of 2–75 µs. The crossover from phase memory to phase reset persists across the guard times examined. **c,** Phase fidelity extracted using Fourier windows spanning three, five and eight oscillation cycles. Horizontal dotted lines in a and c indicate the 1/e fidelity threshold. **d,** Number of valid events $N_{\mathrm{valid}}$ retained for each window length. Three-cycle windows suffer substantial event rejection at longer extinction durations owing to the post-restoration fit quality criterion ($R^2$), whereas five- and eight-cycle windows retain nearly all events and yield closely consistent fidelity trends.

The resulting fidelity curves exhibit consistent behavior across the examined guard times (Supplementary Fig. 3a). Although modest differences occur at intermediate interruption durations, the principal crossover from high phase fidelity to substantial phase displacement remains evident. Similarly, the restoration-phase concentration increases towards unity for longer interruptions, with closely corresponding trends across the guard-time conditions (Supplementary Fig. 3b). These results indicate that the observed transition from phase continuation to reproducible resetting is not strongly dependent on the precise onset of the post-restoration fitting window.

The dependence on fitting-window length was examined using Fourier windows spanning three, five and eight oscillation cycles (Supplementary Fig. 3c). The five- and eight-cycle analyses yield closely consistent phase-fidelity trends, including the characteristic decline across the interruption-duration range. Three-cycle fits exhibit greater variability, particularly at longer interruptions, where the reduced number of oscillation periods limits the stability of phase estimation. This behavior is further reflected in the number of valid events retained after applying the post-restoration fit-quality criterion based on the coefficient of determination, $R^2$ (Supplementary Fig. 3d). Three-cycle windows exhibit substantial event rejection at longer interruption durations, whereas five- and eight-cycle windows retain nearly all recorded events. Accordingly, five-cycle fits were adopted for the primary analysis, providing a balance between temporal localization, phase-estimation stability and event retention. The agreement across the tested analysis conditions supports the robustness of the experimentally observed phase-memory crossover.